\documentstyle[12pt]{article}
\newcommand{\be}{\begin{equation}}
\newcommand{\ee}{\end{equation}}
\newcommand{\bea}{\begin{eqnarray}}
\newcommand{\eea}{\end{eqnarray}}
\newcommand{\delum}{\partial_\mu}
\newcommand{\delom}{\partial^\mu}
\newcommand{\delun}{\partial_\nu}

\newcommand{\komma}{\mbox{ ,}}
\newcommand{\pkt}{\mbox{ .}}
\begin{document}

\title{Stability of the B=2 hedgehog in the Skyrme model}

\author{Thomas Waindzoch\thanks{electronic address:
T.Waindzoch@kfa-juelich.de}\\
\small Institut f\"ur Kernphysik, Forschungszentrum J\"ulich GmbH.\\
\small D-52425 J\"ulich, Germany \\
\\
Jochen Wambach\\
\small University of Illinois at Urbana-Champaign\\
\small Department of Physics, Loomis Laboratory \\
\small 1110 W.Green Street, Urbana IL 61801, USA }
\date{\today}
\maketitle
\noindent

\begin{abstract}
We study the unstable modes of the baryon number two
hedgehog of the Skyrme model on a three dimensional spatial lattice.
An expansion of the Skyrme Lagrangian
around the hedgehog configuration provides the equations of motion
for the fluctuation fields solvable numerically via a
relaxation method.
We find the negative energy modes and, by evolving the
excited hedgehog in time, a breakup into two separated
solitonic configurations is obtained.
Different paths of descent for the receding Skyrmions are presented and
the possibility of
determining the metric structure of the collective-coordinate manifold
is discussed.
\end{abstract}

\section{Introduction}
The Skyrme model \cite{Skyrme62}, as a semiclassical model of {\it QCD}
in the non-perturbative regime, is quite successful in describing
low-energy hadronic physics.
Starting from a classical topological Skyrmion solution, the calculated
single-baryon properties are in qualitative agreement with measured
physical values.
This is, in view of the few parameters of the model, quite remarkable.
A challenging problem is to apply the model to more than one baryon and
to study the interaction of baryons within the picture of interacting
Skyrmions.
Due to the non-linear structure of the model, the equations of motion
for more than one Skyrmion are very hard to solve
and usually one avoids this difficulty by
constructing the multi-Skyrmion configuration out of the
known single fields.
However, the often used 'product ansatz' in the two-Skyrmion case, valid for
large separation, cannot reproduce the axially symmetric minimal energy
configuration and therefore fails to give any medium-range attraction in the
central component of the nucleon-nucleon interaction.
This medium-range attraction was found only after performing a full
lattice calculation \cite{VWWW468,Verbaar195,KopSht87}.
This tells us, that one should not rely on approximations for short distances
of the Skyrmions.
They interact strongly and one has to know the
exact field configuration in order to make valid statements about the
physics.\\

Although some features of the
nucleon-nucleon interaction were obtained within the Skyrme model,
many parts of the potential are still unresolved or missing.
As a consequence, the phase shifts for the interaction are
described fairly inaccurately \cite{Walet}.
This insufficient result is mostly due the lack of knowledge on the
short-distance behavior of the Skyrmion dynamics.
This dynamics contains the necessary information for the
semiclassical quantization of the model and a detailed
understanding of the different properties of
interacting Skyrmions becomes very important.
The Skyrmion-Skyrmion interaction potential has been calculated
previously \cite{WahWam91,WaiWam}, but contributions from  the inertia tensor
of the
Skyrmions remain largely unknown.
The momentum operators in the collective-coordinate quantization scheme are
connected to this
inertia tensor and any description will be incomplete as long as the
Hamiltonian misses proper kinetic energy terms.
The investigation of the explicit form of
the inertia tensor and its dependence on the separation of the
two Skyrmions will therefore be the main purpose of the present study.

To gain more insight into the short-distance properties of the two-Skyrmion
sector, we analyze the stability of the B=2 hedgehog on a three dimensional
spatial lattice. This special Skyrmion configuration is important because it
can serve as an starting point for obtaining many different
two-Skyrmion configurations at finite separation \cite{Manton}.
The hedgehog is a saddle point in the energy surface and
thus unstable to certain excitations.
Since the Skyrmions carry a conserved topological number,
the decay path starting from the excited hedgehog will connect
different Skyrmion configurations in the same topological sector.
Following various paths one should be able to construct the full
Skyrmion-Skyrmion interaction potential and the inertia tensor from the
given information.\\

In this paper we wish to perform the first part of this program, namely the
calculation of
the decay paths for the B=2 hedgehog. Therefore the paper is organized as
follows:
in the next section we briefly introduce the Skyrme model Lagrangian and
discuss
the collective-coordinate Hamiltonian for the baryon number B=2 sector.
After presenting the numerical method, we will investigate the stability of the
B=2 hedgehog and explicitly calculate its unstable modes on the lattice.
By evolving the fields in time, we obtain
different paths of descent from the hedgehog and we will discuss the connection
to the collective-coordinate description.
Conclusions and a short outlook on future work will end the paper.

\section{The Skyrme model}
\subsection{The Lagrange density}
The Skyrme model is an effective chiral theory to model $QCD$ in the
non-perturbative regime. For large values of $N_C$, the number of colors for
quarks and gluons, t'Hooft \cite{tHooft}
and later on Witten \cite{Witten} showed, that $QCD$
reduces to a theory of weakly interacting mesons (and glue balls) where the
baryons emerge as solitonic solutions of the classical field equations.
This large-$N_C$ expansion, unfortunately, has never been worked out
systematically due to enormous algebraic and technical problems for 3+1
dimensions. Hence one has to study models, which attempt to capture the
main features of large-$N_C$ $QCD$. These models should obey
the known low-energy theorems for meson physics dictated by chiral symmetry
and current algebra. One of the simplest models, which fulfills this
requirement, is the Skyrme model. It is widely believed, that it represents
the essential parts of the true large-$N_C$ expansion, if it ever will be
developed from $QCD$.\\
The Lagrange density of the Skyrme model can be written as
\cite{Skyrme62}
\be
{\cal L}  =  {{f_\pi^2}\over 4} Tr[\delum U \delom U] +
{1\over{32 e_s^2}}Tr([U^\dagger \delum U,U^\dagger \delun U])^2
 + {{m_\pi^2 f_\pi^2}\over 2}Tr[U-1] \komma
\ee
where the non-linear pion field
$U$ is a unitary $SU(2)$ matrix, obeying the chiral constraint
\be
U U^\dagger = 1   \pkt
\ee
A possible parameterization of U is
of the form
\be
U=(\sigma + i\vec{\tau} \vec{\pi})/{f_\pi} \pkt
\ee
This shows the connection of the model to an effective meson theory for pions.
The sigma field is not a free field, but depends on the pion field via the
unitarity relation (2).
The parameters of the model are the pion decay constant $f_\pi$, the
physical pion mass $m_\pi$ and the 'Skyrme parameter' $e_s$.\\

In addition to the description of low-energy pion interactions, the model
simultaneously allows for baryons. Due to the fourth-order Skyrme term, which
has the right number of spatial derivatives to obtain localized
configurations in space, the field $U$ can be excited as a soliton, stabilized
by topology. These special excitations, called Skyrmions, are characterized by
a topological current
\be
B^\mu ={1\over{24\pi^2}}\epsilon^{\mu \nu \rho \sigma}
Tr[(U^\dagger \partial_\nu U)(U^\dagger \partial_\rho U)(
U^\dagger \partial_\sigma U)]  \komma
\ee
which can be derived from the mapping properties of the fields.
The time component of this current lead to a conserved topological quantity
\be
B=\int d^3x B^0 = n \komma
\ee
representing the winding number of the field $U$.
Physically this number can be identified with the baryon number
\cite{Witten}.\\

For the topological sector with baryon number
B=1, the description of the physical particles like the nucleon or the
$\Delta$-isobar is in qualitative agreement to experiment.
For a detailed discussion of this sector see \cite{ANW}.

\subsection{The B=2 sector of the model}
The Skyrmion sector with the topological number B=2 is the domain
where one studies the deuteron properties and the nucleon-nucleon interaction.
Different solitonic configurations are
known for B=2 and some of them have been studied previously
\cite{KutPet,VWWW468,KopSht87}.
All classical configurations are characterized by a set
of coordinates which transform one
configuration into an other.
In the semiclassical quantization scheme, these
coordinates become operators and allow for the description of the physical
particles. \\

To perform realistic calculations within the model, however, one should
truncate
the set of coordinates from the beginning. Comparing the energy shifts of the
Skyrmion, one can group the corresponding transformations into two essentially
different sets. Vibrations arising from explicit pionic excitations of the
solitonic background field are of relatively high energy compared with the
zero- or nearly zero modes for a given configuration.
One should treat these vibrations via perturbation theory and, to first order,
simply neglect them.
It remains a problem left for future studies to consider explicitly the
vibrational modes of the Skyrmions in order to perform the renormalization
program. Most important for a first description of the physical processes are
the zero- or nearly zero modes.  To first order in a
perturbation theory of semiclassical quantization, these modes cannot be
neglected in the baryonic sector.\\

Going back to the work of Manton \cite{Manton} on the B=2 sector
the minimal number of coordinates for the description of the Skyrmion-Skyrmion
dynamics is twelve. In general there are three coordinates for
global translations, three for spatial rotations and three for isorotations.
The corresponding transformations are related
to the symmetries of the field $U$ and define true zero modes.
The remaining three coordinates describe the orientation of two individual
B=1 Skyrmions relative to each other.
This orientation is determined by one coordinate for the relative separation
and two angles for the relative isospin rotation. These relative coordinates
are connected to modes, which change the energy. The energy shift is rather
small, however, compared to the total field energy at least for large
distances. For example is the energy difference between the 'donut', which is
the axially-symmetric solution to minimal energy,  and two infinitely
separated Skyrmions only a few percent of the total mass of the donut.
This means, that one should treat
these coordinates like the zero modes in the collective-coordinate approach.
They cannot be neglected like the vibrations. \\

In the collective coordinate approach, the $U$-field will have the form
\be
U(x,t)=A(t)U(D(B(t))x-R_t(t),R(t),C(t))A^\dagger(t) \komma
\ee
where the time dependence is shifted to the coordinates
$R_T,B,A,R$ and $C$, which represent the three translations, three rotations,
three isorotations, the relative separation and the relative isorotation
respectively. The relative coordinates are internal coordinates of the field
since they change explicitly the form and the energy of the soliton.
Using this ansatz one arrives at the following
classical collective-coordinate Hamiltonian for the B=2 sector of the
Skyrme model
\be
H={1\over 2}P_i M_{ij}^{-1}(R,C) P_j + U(R,C) + 2M_H \label{hamiltonian}  \pkt
\ee
Here $P_i = M_{i,j}(R,C) \dot{Q_j}$ are the canonical momenta
conjugate to the coordinates
\be
Q_i=(A,B,R_T,R,C) \pkt
\ee
The Skyrmion-Skyrmion interaction potential $U(R,C)$ and the mass matrix
$M_{ij}(R,C)$ are functions of the relative coordinates R and C.
For quantization, the collective coordinates and their momenta are replaced
by operators which fulfill the canonical commutation relations
\be
[\hat{P_i},\hat{Q_j}]=i\delta_{ij}  \pkt
\ee
This will lead to a quantum Hamiltonian for the B=2 sector
capable of describing the low-energy interaction of two baryons. \\

Before one can go over to the study of physical processes one first has
to reliably determine the entries of the classical Hamiltonian.
The interaction potential $U(R,C)$ has been determined previously on a
spatial lattice \cite{WahWam92}. One is then left with the
evaluation of the mass matrix $M_{ij}(R,C)$. One method is to use the explicit
dependence of the fields on the collective coordinates.
This leads to the expression
\be
M_{ij}(R,C) = \int d^3x {{\partial \phi_\alpha}\over{\partial Q_i}}
M_{\alpha\beta}{{\partial \phi_\beta}\over{\partial Q_j}}  \label{mass}
\ee
where $M_{\alpha\beta}$ is the metric of the Skyrme model Lagrangian. Its form
is given in
(\ref{metric}).
Since the B=2 configurations are changing under the relative transformations,
it implies that one has to calculate the fields at many different points
$(R,C)$ in order to determine the field derivatives to the coordinates $Q_i$.
This is very tedious and some results will be presented in a forthcoming paper
\cite{WaiWam2}. \\

Alternatively, one can perform time-dependent simulations of the dynamics.
Starting from some initial condition, the solitonic motion can be followed
numerically and by comparing the results with paths from the collective
coordinate Hamiltonian (\ref{hamiltonian}) one should be able to deduce the
structure of the
mass matrix. In terms of the collective coordinates, the equations of motion
for the time evolution are of the form
\be
M_{ij}(R,C) \ddot Q_j + \Gamma_{ijk}^{(1)}(R,C) \dot{Q_j} \dot{Q_k}
= -{{\partial U(R,C)}\over {\partial Q_i}}  \label{hameq}
\ee
where the Christoffel symbols of the first kind are
\be
\Gamma_{ijk}^{(1)}(R,C)={1\over 2}[
{{\partial M_{ij}(R,C)}\over {\partial Q_k}}
+{{\partial M_{ik}(R,C)}\over {\partial Q_j}}
-{{\partial M_{jk}(R,C)}\over {\partial Q_i}} ] \pkt
\ee
If the collective-coordinate picture is valid, these equations of motion
should describe the motion of the soliton fields and therefore
be directly comparable to numerical lattice results.
By calculating the relative coordinates and the potentials for a given path
$\Gamma=(R,C)$ in the simulation one could therefore determine the mass matrix.
In the following we will see to what degree this will be possible. Since
our interest is in the short-distance dynamics of the two-Skyrmion system we
will start from an initial B=2 hedgehog.

\section{The numerical method}
Before studying the stability of the B=2 hedgehog on the lattice
we wish to discuss the numerical method used in this paper.
To perform calculations for the field $U$ on a three-dimensional lattice
we switch to the parameterization
\be
U=\phi_0 +i\vec{\tau}\vec{\phi} \komma
\ee
where the four fields $\phi_\alpha $ are now constrained to lie on the unit
chiral circle
\be
\phi^2_{\alpha}=1  \pkt
\ee
Here the greek indices run from 0 to 3 while the latin indices run from 1 to 3.
The constraint has to be fixed via a Lagrange multiplier
$\lambda$ such that the Lagrange density becomes
\bea
{\cal L}&=& {{f_\pi^2}\over 2}(\delum\phi_\alpha)^2
-{1\over{4e_s^2}}(\delum\phi_\alpha)^2(\delun\phi_\beta)^2
+{1\over{4e_s^2}}(\delum\phi_\alpha\delum\phi_\beta)^2 \nonumber\\
&& + m_\pi^2 f_\pi^2(\phi_0-1) + {1\over 2}\lambda(\phi_\alpha^2 -1) \pkt
\eea
Since we are interested here in the classical dynamics
the Euler-Lagrange equations for the fields
$\phi_\alpha$
\be
\delum{{\delta {\cal L}}\over {\delta \delum \phi_\alpha}}
-{ {\delta {\cal L}}\over {\delta \phi_\alpha}} = 0
\ee
have to be solved. Using the canonical momenta
$\pi_\alpha = {{\delta {\cal L}}\over {\delta \partial_t \phi_\alpha}}$
to the fields $\phi_\alpha$, one obtains a set of coupled non-linear
partial differential equations
\bea
\partial_t \phi_\alpha &=& {M_{\alpha\beta}^{-1}}\pi_\beta \komma \\
\partial_t \pi_\alpha &=& \partial_i (C_{\alpha\beta}\partial_i \phi_\beta)
+m_\pi^2 f_\pi^2 \delta_{0 \alpha} +\lambda \phi_\alpha    \pkt
\eea
The $4\times4$ matrices $M_{\alpha\beta}$ and $C_{\alpha\beta}$ are functionals
of the fields and have the form
\bea
M_{\alpha\beta}&=& [f_\pi^2 + {1\over{e_s^2}}(\partial_i \phi_\gamma)^2]
\delta_{\alpha\beta} - {1\over{e_s^2}}\partial_i \phi_\alpha \partial_i
\phi_\beta
\label{metric} \komma\\
C_{\alpha\beta}&=& [f_\pi^2 - {1\over{e_s^2}}(\partial_\mu \phi_\gamma)^2]
\delta_{\alpha\beta} + {1\over{e_s^2}}\partial_\mu \phi_\alpha \partial_\mu
\phi_\beta
\eea
where $M_{\alpha\beta}$ is the metric of the Skyrme model.

To solve these equations numerically a finite differencing scheme is employed.
The fields and the momenta are defined on a cartesian lattice $(i,j,k)$ and
the spatial derivatives are evaluated via central differences as
\bea
\partial_x \phi_\alpha (i,j,k) & = &
{{\phi_\alpha (i+1,j,k) -\phi_\alpha (i-1,j,k)}\over {2\Delta x}} \komma \\
\partial_x^2 \phi_\alpha (i,j,k) &=&
{{\phi_\alpha (i+1,j,k) -2\phi_\alpha (i,j,k) + \phi_\alpha (i-1,j,k)}
\over {{\Delta x}^2}}
\eea
and so on.\\
For the time evolution we use the Leapfrog method
\bea
\pi_\alpha^n & = & \pi_\alpha^{n-2} + 2\Delta t [
\partial_i (C_{\alpha \beta}^{n-1} \partial_i \phi_\beta^{n-1})
+m_\pi^2 f_\pi^2 \delta_{0 \alpha} +\lambda \phi_\alpha^{n-1}] \komma \\
\phi_\alpha^{n+1} &=& \phi_\alpha^{n-1} + 2\Delta t [
(M_{\alpha \beta}^{-1})^n \pi_\beta^n ] \pkt
\eea
As initial conditions the fields have to be predetermined at the two
time steps $t=-1,0$ and the momenta at the time steps $t=-2,-1$.
The fields and momenta on the lattice boundary are set to their
vacuum values.
The Lagrange parameter $\lambda$ is adjusted during the iteration process
to ensure the chiral constraints $(\phi_\alpha^{n+1})^2 =1$ and
$\phi_\alpha^n \partial_t \phi_\alpha^n = 0.$
The latter results from the time derivative of the first  and
both have to be satisfied simultaneously.\\

The Leapfrog method is easily adapted as a relaxation method for
calculating static solutions to the field equations.
Considering time as pseudotime, one
performs the calculation of the momenta at pseudotime step $t^n$ with
vanishing momenta and time derivatives from previous pseudotime steps.
This method advances the fields to a static configuration since
the configuration loses more and more of its kinetic energy as pseudotime
progresses.\\

In this paper the lattice is kept rather small in order to limit the amount
of computational time. We take 42 points for each spatial
direction with a grid distance of $\Delta x = 0.06 fm$. The physical volume of
our 'cube' is thus $V=(2.8 fm)^3$. This size is large enough for a calculation
of the unstable modes of the B=2 hedgehog but is certainly to small to
simulate the time evolution as the hedgehog breaks up and individual
Skyrmions emerge.
For numerical stability we have to ensure the Courant-Levi condition
\be
{{v \Delta t}\over {\Delta x}} \ll 1 \komma
\ee
where $v$ is a velocity scale for the evolution of a given field configuration
($v\approx c$). Typically this ratio is around 0.1 in our calculations.
There are, however, additional stability problems which will be discussed
later in the paper.\\

The parameters of the Skyrme model are set to $f_\pi = 93 MeV,m_\pi=138 MeV$
and $e_s = 4.76$. This choice ensures the correct strength of the asymptotic
one-pion exchange potential for the nucleon-nucleon interaction within the
model. However, the mass of a B=1 Skyrmion, $M_{B=1}=1416 MeV$, turns out too
large. This should be compared to the continuum limit of $M_{B=1}=1463 MeV$.
A lattice error of about $5\%$ is tolerable for a first study.

\section{Stability of the B=2 hedgehog on the lattice}
\subsection{The unstable manifold of Manton}
In the language of geometry the mass matrix $M_{ij}(R,C)$ is nothing but
the metric of the twelve dimensional collective-coordinate manifold for the
B=2 sector. In contrast to interacting magnetic monopoles \cite{ManGib},
the motion on this manifold is not geodesic due to the presence of the
potential energy $U(R,C)$ (see eq.~(\ref{hameq})). Following different paths on
 the
collective subspace of Skyrmion configurations one can thus draw conclusions
to the geometrical and potential structure on the manifold.
As a source of paths, Manton proposed to consider the B=2 hedgehog in
more detail. This special configuration is not stable, but represents a
saddle point in the energy plane. It has six different zero modes rather than
nine for the most general B=2 configuration. The hedgehog symmetry simply
connects space to isospin space and global rotations become equivalent to
global isorotations, thus reducing the number of degrees of freedom by three.
The B=2 hedgehog has also six unstable modes. This was already discussed by
Manton and the modes were explicitly found later for the $S^3$-sphere
\cite{WiBa}. These modes lower the energy of the hedgehog and define the
starting points for paths of steepest descent. Together with the zero modes
one has, by superposition of unstable modes, a large number of available
initial configurations to probe the manifold.\\

\subsection{The B=2 hedgehog on the lattice}
The hedgehog is determined by the ansatz
\be
U(x)=\cos{F(r)} + i\vec{\tau}\hat{x}\sin{F(r)}
\ee
where the boundary conditions for the profile function $F(r)$ are chosen to
ensure baryon number B=2:
\bea
F(0)=2\pi \komma F(\infty)=0 \pkt
\eea
One can easily obtain the profile function by minimizing the static energy for
this ansatz numerically by solving the radial differential equation.
For our parameter set we calculate the B=2 hedgehog
mass as $M_2=4418 MeV$. This is nearly three times the B=1 hedgehog mass of
$M_1=1463 MeV$.\\

For a numerical stability analysis of the B=2 hedgehog the continuum result is
unsuited,
however. Rather one has to start from the minimized lattice hedgehog.
Thus we use the free hedgehog profile function to set up an initial
configuration which is then relaxed via the procedure introduced above.
To improve convergence we reduce the cube to one octant
via boundary conditions given by the symmetries of the hedgehog ansatz.
After several thousand iteration steps we obtain a lattice hedgehog mass
of $M=4118 MeV$ and a baryon number of $B=1.83$. A comparison of the free
profile functions with the extracted lattice profile function is shown in
fig.~1 .
\begin{figure}[h]
\vskip 9.5cm
\includegraphics{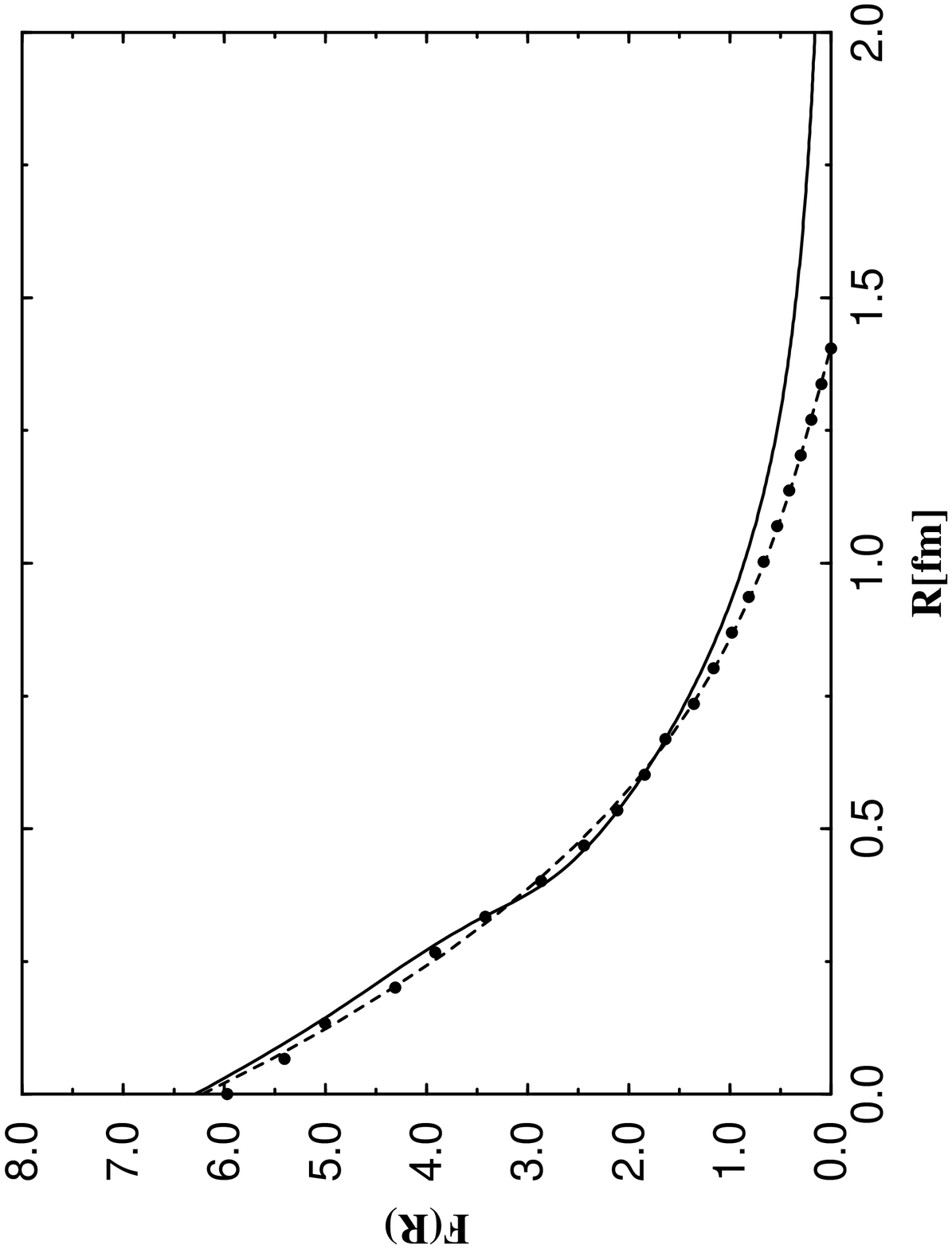}
{\small Fig.~1. The free (solid)
and lattice (dots) profile function for the B=2 hedgehog.
\par}
\end{figure}

To check the long-time stability of the lattice hedgehog we have evolved the
fields in real time via the leapfrog method. Even after several thousand time
steps no appreciable gain in kinetic energy has been detected.

The unstable modes of the hedgehog are modes of lower energy and
are calculable from an expansion of the Lagrange density around the background
fields $\phi_\alpha^H$. Inserting the ansatz
\be
\phi_\alpha = \phi_\alpha^H +\delta \phi_\alpha \label{fullconf}
\ee
into the Skyrme Lagrangian and keeping only terms up to second order in
the small-amplitude fluctuations $\delta \phi_\alpha$, one obtains the
expansion
\be
{\cal L}[\phi]\approx {\cal L}[\phi^H]+{\cal L}^{(2)}[\phi^H,\delta\phi] \pkt
\ee
Terms of first order in the fluctuations have to vanish due to the equations
of motion for the extremal background field $\phi_\alpha^H$.\\

The explicit form of the second-order Lagrange density is
\be
{\cal L}^{(2)}[\phi^H,\delta\phi] =
{\cal T}^{(2)}[\phi^H,\delta\phi]
-{\cal V}^{(2)}[\phi^H,\delta\phi] \komma
\ee
with the kinetic and potential energy densities given by
\bea
{\cal T}^{(2)}[\phi^H,\delta\phi]&=&
{1\over 2}(\partial_t \delta\phi_\alpha)M_{\alpha\beta}[\phi^H]
(\partial_t \delta\phi_\beta) \komma \\
{\cal V}^{(2)}[\phi^H,\delta\phi]&=&{{f_\pi^2}\over 2}(\partial_i
\delta\phi_\alpha)^2
+{1\over {2 e_s^2}}[
(\partial_i \phi_\alpha)^2(\partial_j \delta \phi_\beta)^2
+2(\partial_i\phi_\alpha \partial_i\delta\phi_\alpha)(\partial_j\phi_    \beta
\partial_j \delta\phi_\beta) \nonumber \\
&&- (\partial_i\phi_\alpha \partial_i\phi_\beta)
(\partial_j\delta\phi_\alpha \partial_j \delta\phi_\beta)
-(\partial_i\phi_\alpha \partial_j\phi_\alpha)
(\partial_i\delta\phi_\beta
\partial_j \delta\phi_\beta)  \nonumber \\
& &-(\partial_i\phi_\alpha \partial_j\delta\phi_\alpha)(\partial_j\phi_\beta
\partial_i \delta\phi_\beta)]
 - {1\over 2}\lambda (\delta\phi_\alpha)^2
\eea
Here $M_{\alpha \beta}[\phi^H]$ again denotes the Skyrme model metric.
The parameter $\lambda$ ensures the chiral constraint and can be deduced from
its equation of motion.\\

We use the leapfrog method for the second-order Lagrangian in order to
determine the low-energy part of the mode spectrum. Starting from some initial
fluctuation we relax the system to minimal energy
\be
E^{(2)}=\int d^3x {\cal V}^{(2)}[\phi,\delta\phi]
\ee
until a stable solution $\delta\phi_\alpha(x)$ is found.
During the iteration process we have to ensure the chiral constraint
\be
\phi_\alpha^H \delta\phi_\alpha=0
\ee
via the usual Lagrange multiplier technique. The metric $M_{\alpha\beta}$ is
used as an integration measure providing the normalization condition
\be
\int d^3x \delta\phi_\alpha M_{\alpha\beta}[\phi^H]\delta\phi_\beta =
{1\over{e_s^3f_\pi}} \pkt
\ee
which normalizes the fluctuation fields to unity in dimensionless units.\\

Having found the first unstable mode,
we then choose a new initial field and repeat the iteration scheme, while
keeping the new field orthogonal (within the metric) to the previously
determined fluctuation. In this way we can calculate a number of modes for
a given background field. Unfortunately, our numerical method is only capable
to determine locally stable configurations rather then absolute minima.
We can easily miss some modes and the mode spectrum could be incomplete.
However, a detailed analysis of the full spectrum is not the intention of this
paper.\\

Following this scheme we have determined several negative-energy modes.
Starting from an initial field configuration perpendicular to the
hedgehog and with vanishing $\phi_0$ component, the energy of the most
unstable mode is found in our normalization to be
\be
E^{(2)}_1=-12.3 MeV
\ee
The fluctuation field can be parameterized by a single function $g(r)$ and is
of the form
%\bea
%\delta\phi_0 &=& 0 \komma \nonumber\\
%\delta\phi_1 &=& -g(r)\hat{y}\komma \nonumber\\
%\delta\phi_2 &=& g(r)\hat{x} \komma \nonumber\\
%\delta\phi_3 &=& 0 \pkt \\
%\eea
\be
\delta\phi_0 = 0 \komma\,\,
\delta\phi_1 = -g(r)\hat{y}\komma\,\,
\delta\phi_2 = g(r)\hat{x} \komma\,\,
\delta\phi_3 = 0 \pkt \\
\ee
Considering the spatial permutation symmetry of this solution one
can obtain two other modes with the same energy. The degeneracy is
therefore three. A comparison of the lattice fluctuation mode and the
most unstable 'magnetic' mode on the $S^3$-sphere, calculated by Wirzba et al.
\cite{WiBa}, is displayed in fig.~2.\\
\begin{figure}[h]
\vskip 9.5cm
\includegraphics{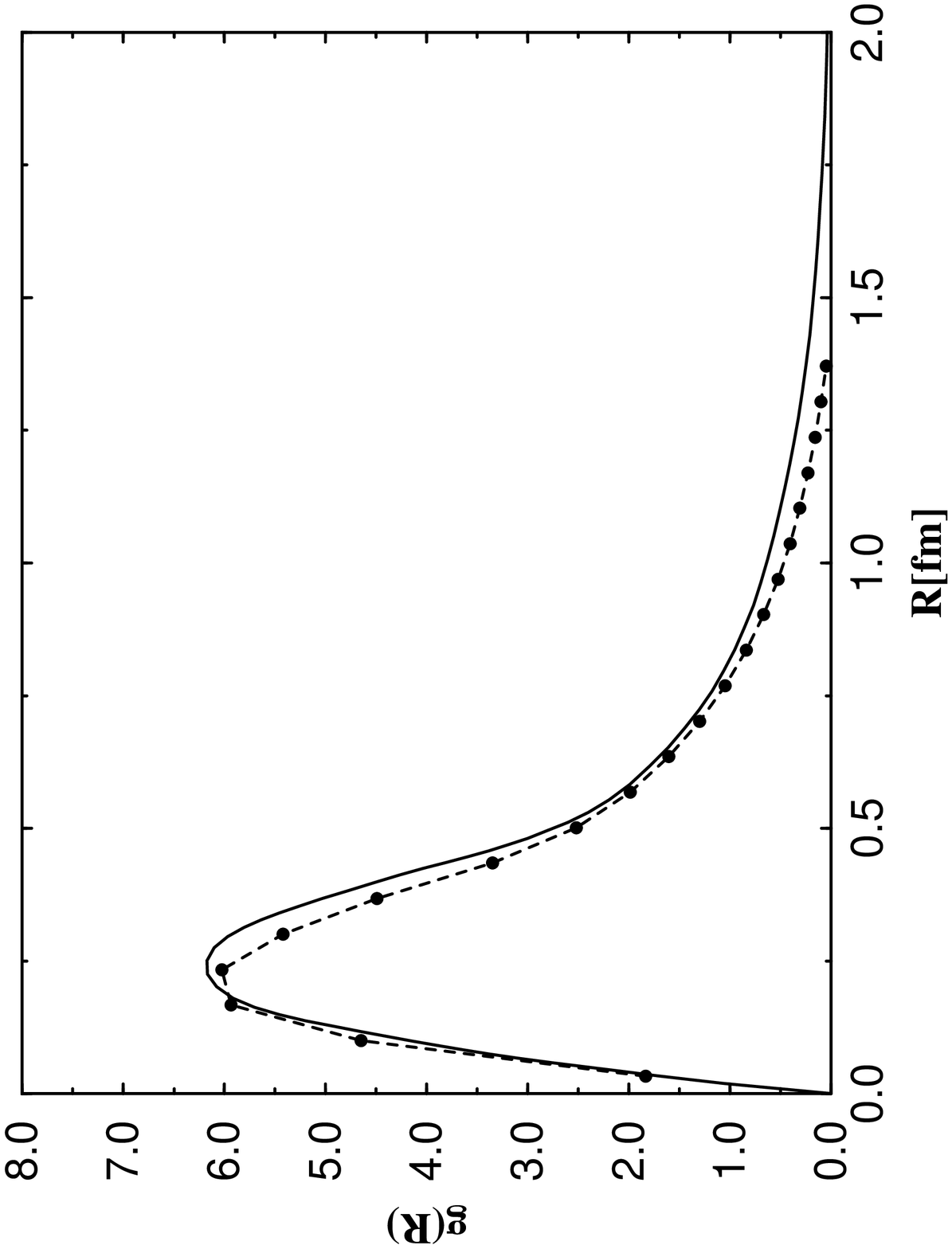}
{\small Fig.~2. The unstable 'magnetic' mode of the B=2 hedgehog.
The solid line denotes the $S^3$-sphere result for vanishing pion mass while
the dots give the result for the lattice.
\par}
\end{figure}
Although the parameter set used by Wirzba et al. (vanishing pion mass)
is slightly different to ours, the agreement is almost perfect.
Assuming a sinusoidal time dependence for the fluctuation field, the
fluctuation energy can also be expressed as imaginary frequency via
the relation \cite{WiBa}
\be
\omega_1^2=(2e_s^3f_\pi)E \pkt
\ee
This yields $\omega_1 = -i 497 MeV$,
to be compared to the $S^3$ result of $\omega_m= -i 460 MeV$.\\

The next higher-lying mode is found by choosing a non-vanishing zero component
for the initial fluctuation field. After relaxation, we obtain the energy of
the new field as
\be
E^{(2)}_2=-6.4 MeV \komma
\ee
with a imaginary frequency of $\omega_2 = -i 358 MeV $.
This results can be compared with the 'electric' mode, the next unstable mode
of the hedgehog on the $S^3$-sphere with a frequency $\omega_e=-i260MeV$.
Since the comparison of the spatial form of the electric fields
is almost as good as for the magnetic mode we believe that the major
difference between the frequencies is due to lattice errors.
The next higher (degenerate) mode found on the lattice has energy:
\be
E^{(2)}_3= -1.9 MeV
\ee
and the spatial form is very close to the zero modes of the B=2 hedgehog
in the continuum. A similar lattice analysis can be performed for
the B=1 hedgehog. The lowest-energy mode is found to be
\be
E^{(2)}_{B=1} = -0.7 MeV \komma
\ee
which is of the same order of magnitude (compared to the total hedgehog
mass) as in the B=2 case. Since the B=1 hedgehog is absolutely stable, this
fluctuation mode should be a true zero mode and indeed, by comparison, the
form of the lattice fields is very similar to the continuum fields.
Assuming that we can correct for lattice errors by shifting the
entire
spectrum such that the B=2 'zero mode' (40) has zero energy, we obtain
\bea
\omega_1&=&-i457MeV \komma\\
\omega_2&=&-i300MeV
\eea
for the two different unstable modes of the lattice hedgehog.
These values are now very close to the $S^3$-sphere results.
The finite pion mass seems to have only minor influence on the
B=2 hedgehog stability. To see, whether the calculated fluctuations really
represent the unstable modes, we should look explicitly to the time
evolution because imaginary frequencies imply exponential growth of the
instability. This is studied in the next section.
\section{The paths of descent for the B=2 hedgehog}
\subsection{The magnetic mode}
With the calculated lattice fluctuation fields we can excite the
B=2 hedgehog in a desired direction to run down the path of
steepest descent. Via superposition of the background field with the
calculated unstable modes and evolving the system in time we should find
different splitting modes for the B=2 hedgehog. Starting first with the most
unstable magnetic mode, a contour plot of the baryon number density at a
fixed value of 0.2 and different time snapshots is shown in fig.~3.
\begin{figure}[h]
\vskip 9.5cm
\includegraphics{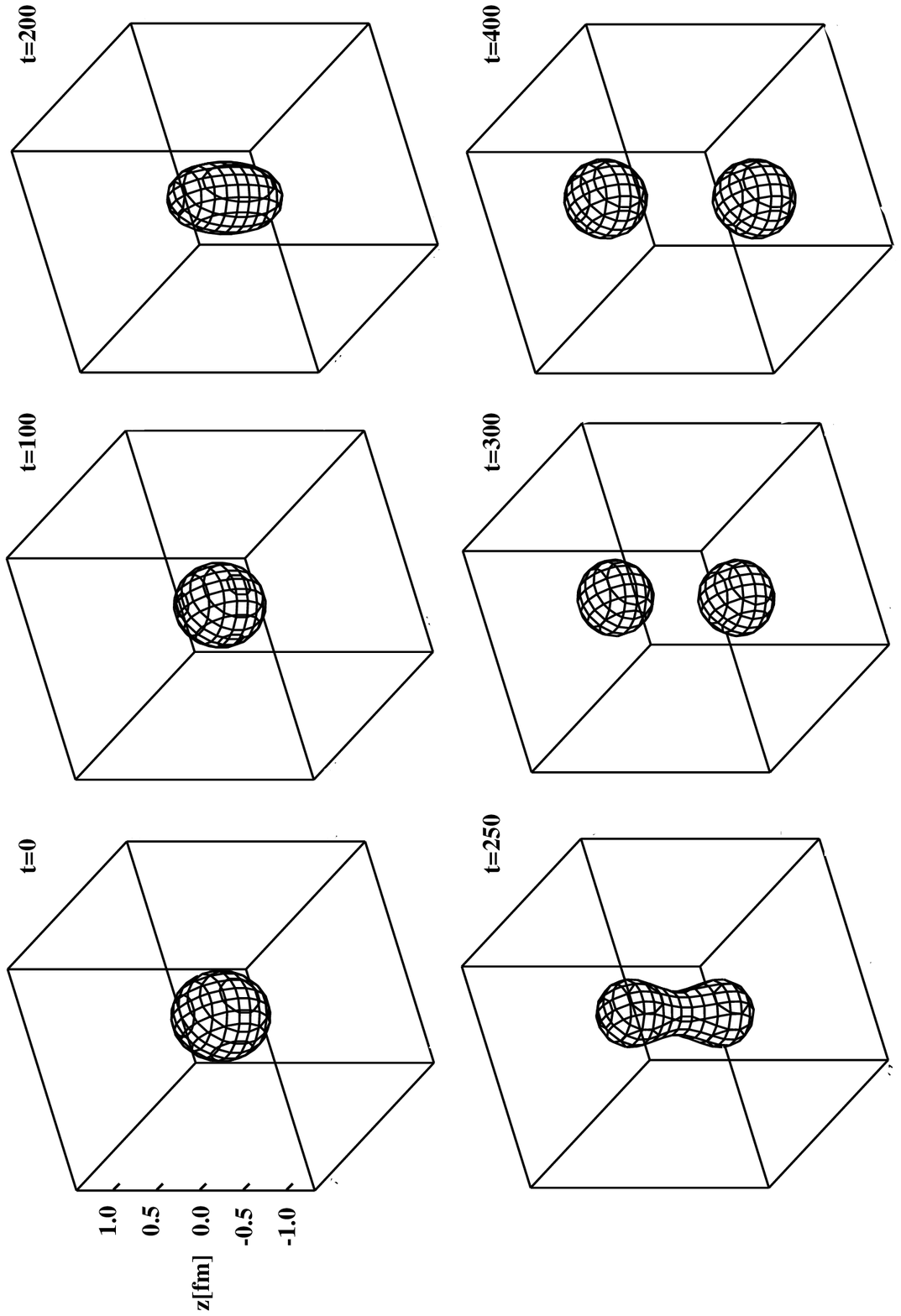}
{\small Fig.~3. The path of descent for different times (in units of the
time step $\Delta t$) after excitation of the magnetic mode of the B=2
hedgehog.
Shown is the contour plot of the baryon number density for the value
$B^{(0)}(x)=0.2$.
\par}
\end{figure}
In the beginning the hedgehog hardly moves. The system is only very slowly
responding to the fluctuation. After a certain amount of time, however,
the configuration rapidly starts to stretch in one direction and finally
splits into two -- almost radially symmetric -- B=1 Skyrmions. In analogy
to nuclear physics this is the 'fission' mode for the B=2 hedgehog. A
different view of the breakup process is given in fig.~4,
where we have plotted the time evolution of a cut through the baryon number
density along  the direction of separation (the z axis).\\
\begin{figure}[h]
\vskip 9.5cm
\includegraphics{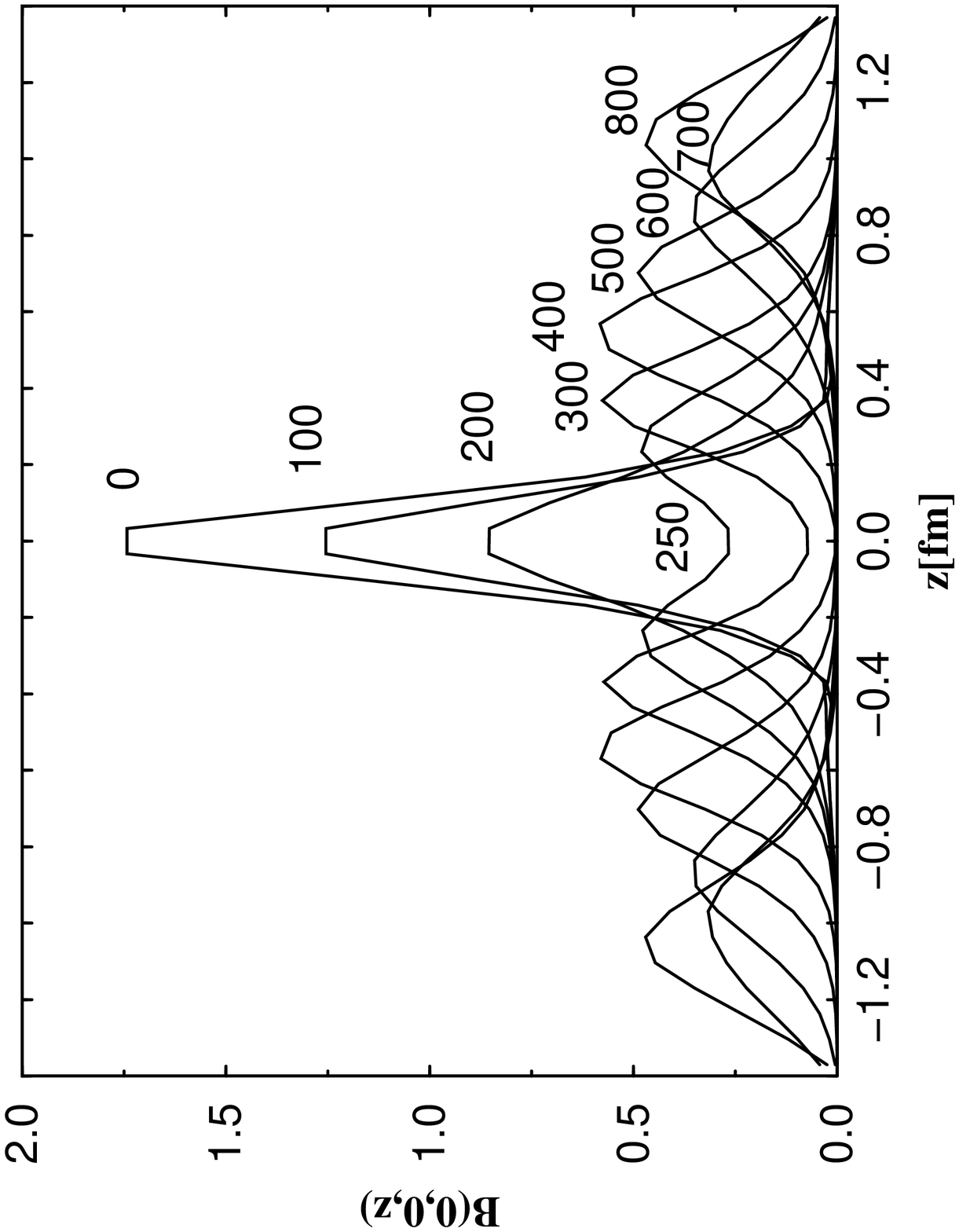}
{\small Fig.~4. The time evolution of a cut through the baryon number density
along the z-axis (axis of separation).
\par}
\end{figure}

In order to numerically follow the path over long time periods we have
to damp the motion along the path. Due to the high energy of the B=2 hedgehog
configuration as compared to twice the B=1 hedgehog mass the outcoming
Skyrmions carry more than $1 GeV$ in kinetic and potential energy in addition
to their rest masses. This energy, initially stored in the B=2 hedgehog, is
gradually converted into deformation energy before it is released over a short
time period in the fission process. At the fission point the kinetic energy
density builds up rapidly and sharp peaks evolve which lead to a complete
breakdown of the numerical method. The local field variations are
simply to rapid to be correctly treated by the leapfrog method.
However, the instability might have a deeper reason connected to a change of
the characteristic of the Skyrme model equation of motion as discussed by
Crutchfield et al. \cite{Crutch}. Due to the high kinetic energy of the fields
at some points on the lattice the hyperbolic equation of motion could change
into a non-hyperbolic differential equation. The latter type is very hard to
control and is simply not tractable within our method.
This problem needs much more advanced algorithms than we have at our
disposal.\\

In order to cross the critical point we introduce an artificial viscosity in
the equation of motion, as  was done in the calculation of the
scattering of two Skyrmions \cite {VWWW461}. This viscosity dissipates
kinetic energy,
but it should not change the paths too drastically since it acts isotropically.
To keep the influence on the dynamics minimal, the viscosity is turned on
as late as possible and is switched off shortly after the critical point.
Our numerical method remains stable, even with more than $700 MeV$ kinetic
energy. This shows that only rapid local changes of the fields
are critical, not large kinetic energies in general. \\

The total energy and the contributions from  kinetic and potential energy
for the fission mode is shown in fig.~5.
\begin{figure}[h]
\vskip 9.5cm
\includegraphics{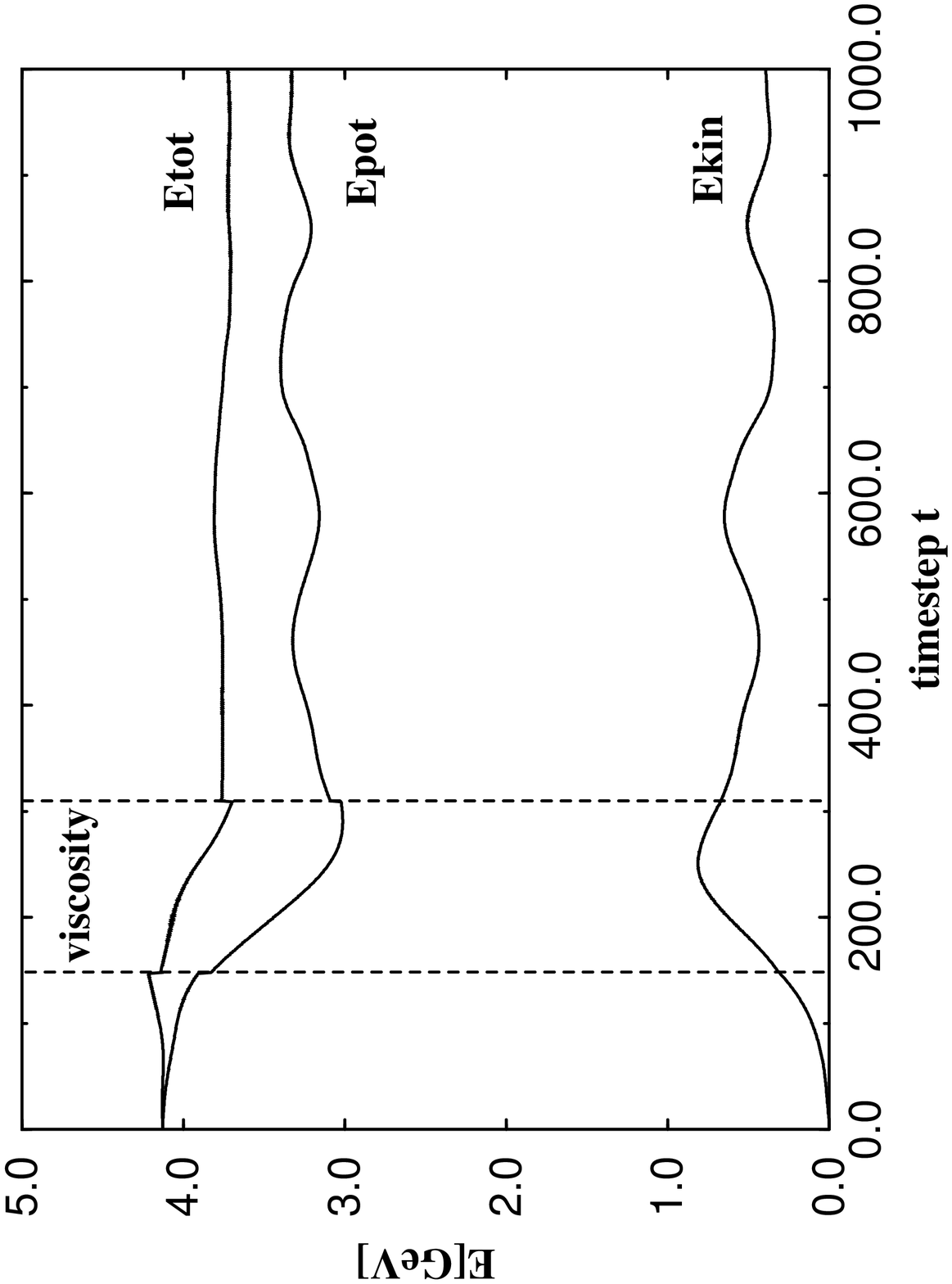}
{\small Fig.~5. The energies for the path of descent for the fission
mode. The viscosity is turned on at times between the two dashed lines.
\par}
\end{figure}
Starting from zero kinetic energy, the B=2 solution gains very little
kinetic energy in the beginning. Suddenly, the Skyrmions roll down the
potential hill away from the B=2 hedgehog saddle and the full deformation
energy of the hedgehog is transferred into kinetic energy of the receding
Skyrmions. The viscous slow down leads to a decrease in total energy but,
after the viscosity is turned off, the total energy remains almost constant.
At that time an almost periodic exchange of potential and kinetic energy takes
place partly hinting towards excitation of vibrational modes (see below).
Such behavior has been observed previously in the numerical study of Skyrmion
collisions \cite{VWWW461}.\\

In addition we have calculated the relative coordinates $R$ and $C$ as
time progresses. The relative separation is defined as the baryonic rms
radius
\be
R=(\int d^3 x B^0(x){\vec{r}\,\,}^2)^{1\over2}
\ee
while the relative isospin orientation can be determined via the symmetry
relation
\be
U(x,C)=C(\vec{\tau}\vec{n})U(D(i\vec{\tau}\vec{n})x,C)
\vec{\tau}\vec{n}C^{\dagger}
\label{sym}
\ee
where $\vec{n}$ is a vector orthogonal to the axis of separation.
Choosing the separation axis as the z-axis, relation (\ref{sym}) connects
the fields for $z>0$ to the fields at $z<0$ and can be used
to compute the relative isospin angles.

\begin{figure}[h]
\vskip 7.5cm
\includegraphics{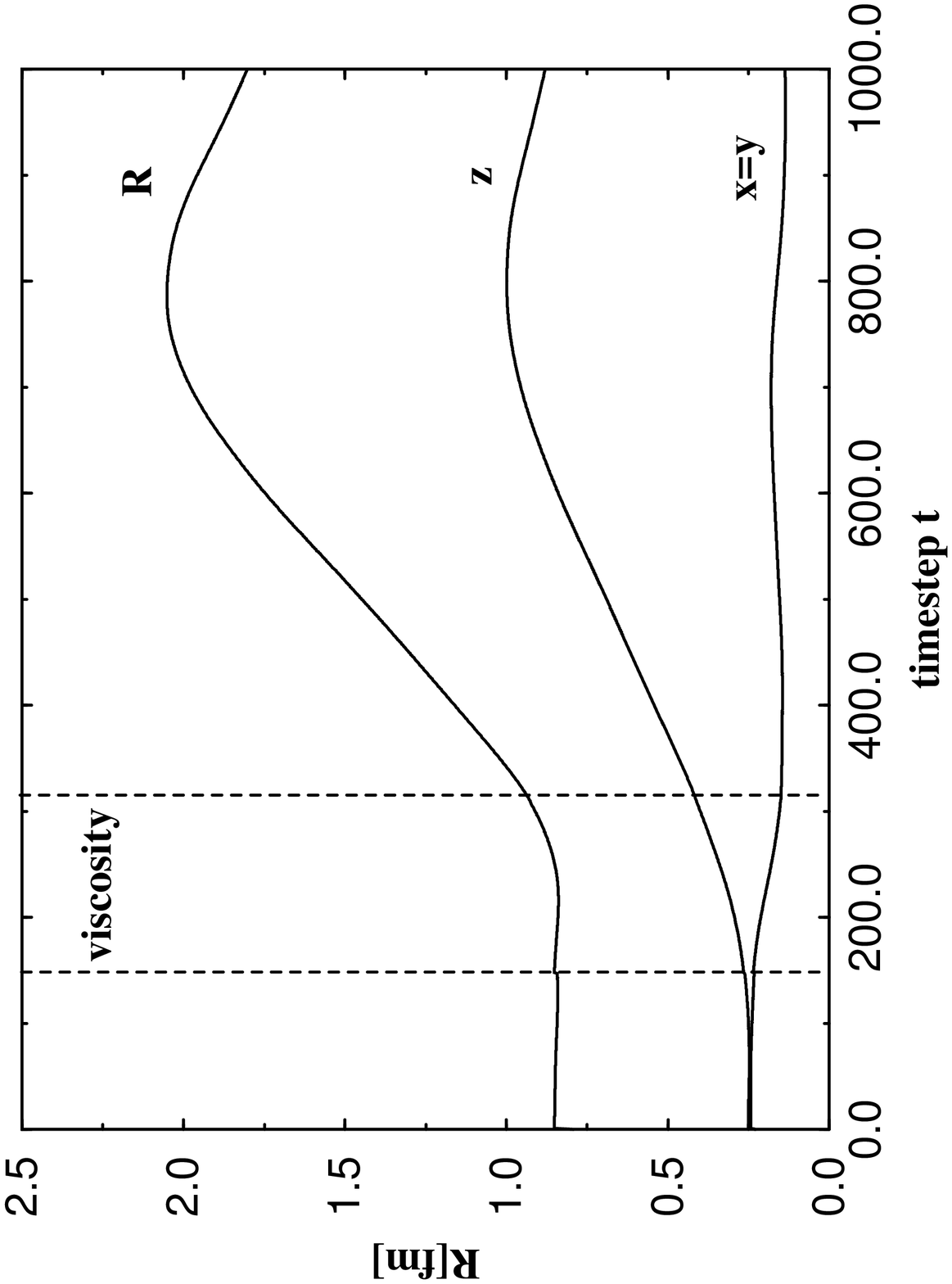}
\includegraphics{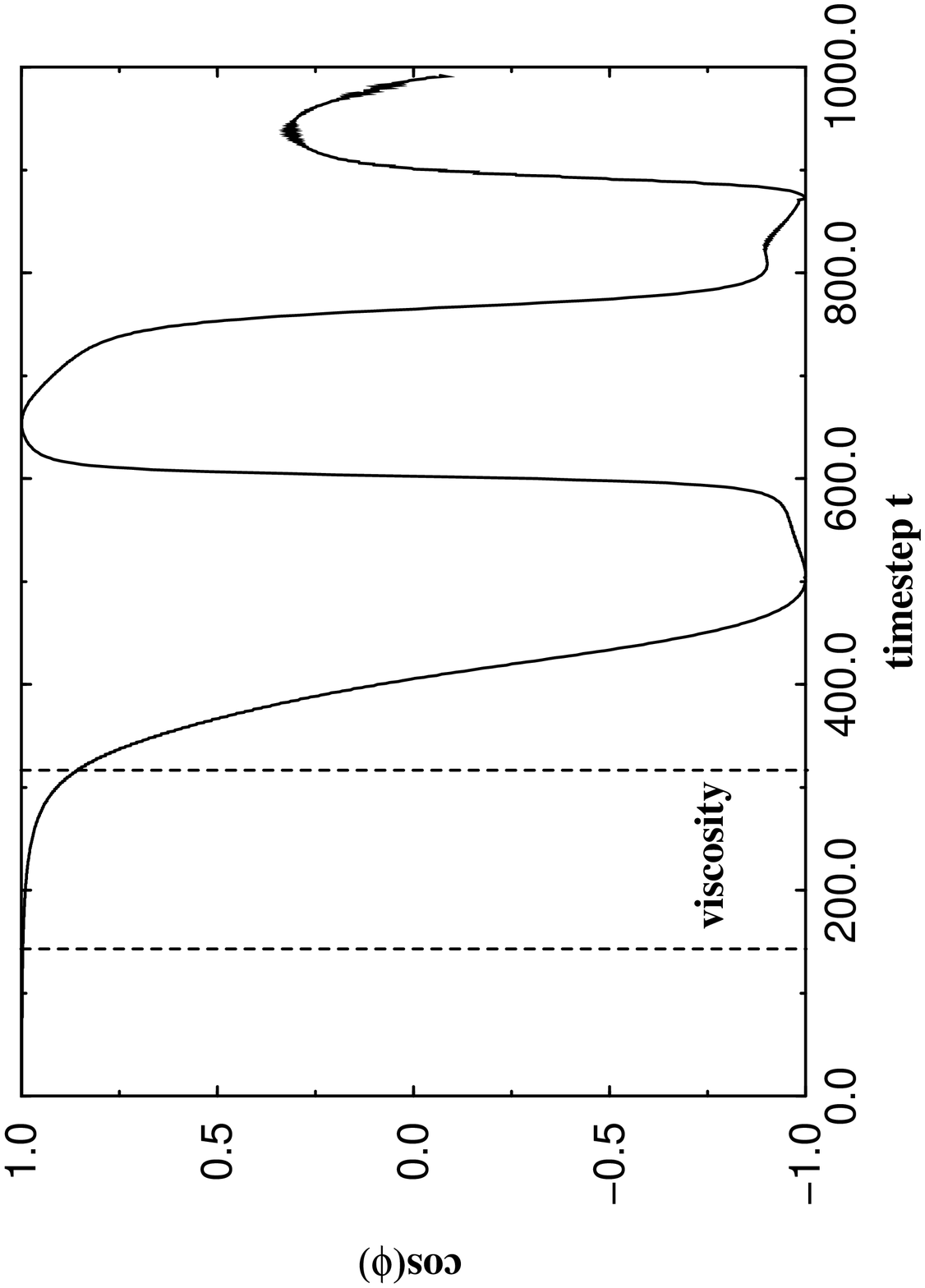}
{\small Fig. 6. The relative coordinates for the path of descent for the
fission mode.
\par}
\end{figure}
In fig.~6 the time dependence of both $R$ and $C$,
parameterized in the form
\be
C=\cos{\phi}+i\tau_3\sin{\phi},
\ee
are given. The latter shows that the receding Skyrmions, which separate in
the z-direction for this magnetic mode, rotate relative to each other around
the separation axis. This rotation in isospin space can explain part of the
periodic exchange of potential and kinetic energy, since the rotation in the
relative coordinate space is connected to a energy shift of the Skyrmions.
At late times the  repulsive lattice boundary starts to become important.
It slows down the Skyrmions and eventually scatters them back towards the
center, as can be seen from the time dependence of $R$ in the left part of
fig.~6. To decrease the effect of the boundary and
allow for undisturbed free motion of the
two skyrmions at larger separations we would have to use a much larger
lattice volume.
The dominant part of the dynamics, however, arises in our opinion
from the intrinsic deformation energies of the receding Skyrmions.
In the beginning the gain in kinetic energy is only due to the new orientation
of the fields to form the separated two single Skyrmions. The motion in
relative coordinate space is there rather slow. Subsequently, as can best be
seen in fig.~4, the single Skyrmions become strongly deformed. They exhibit
a large compression shortly after the critical point and vibrate.
This vibration is responsible for a considerable potential energy at large
separation where one would expect a rather small potential energy from
relative motion. The strong excitation of vibrational modes after the fission
point implies that the collective-coordinate description in terms of zero
and near zero modes along the path of descent is not well justified.
\subsection{The electric mode}
We have performed the same calculation for the less negative unstable mode
which can be identified with the 'electric' mode of Wirzba et al. \cite{WiBa}.
Also in this case viscosity has to be included to cross the critical point
of the evolution. The critical point occurs at the same timestep as in
the previous case. In order to visualize the breakup process for
the 'electric' mode we display the baryon number density for two directions
as a surface plot in fig.~7.
\begin{figure}[h]
\vskip 9.5cm
\includegraphics{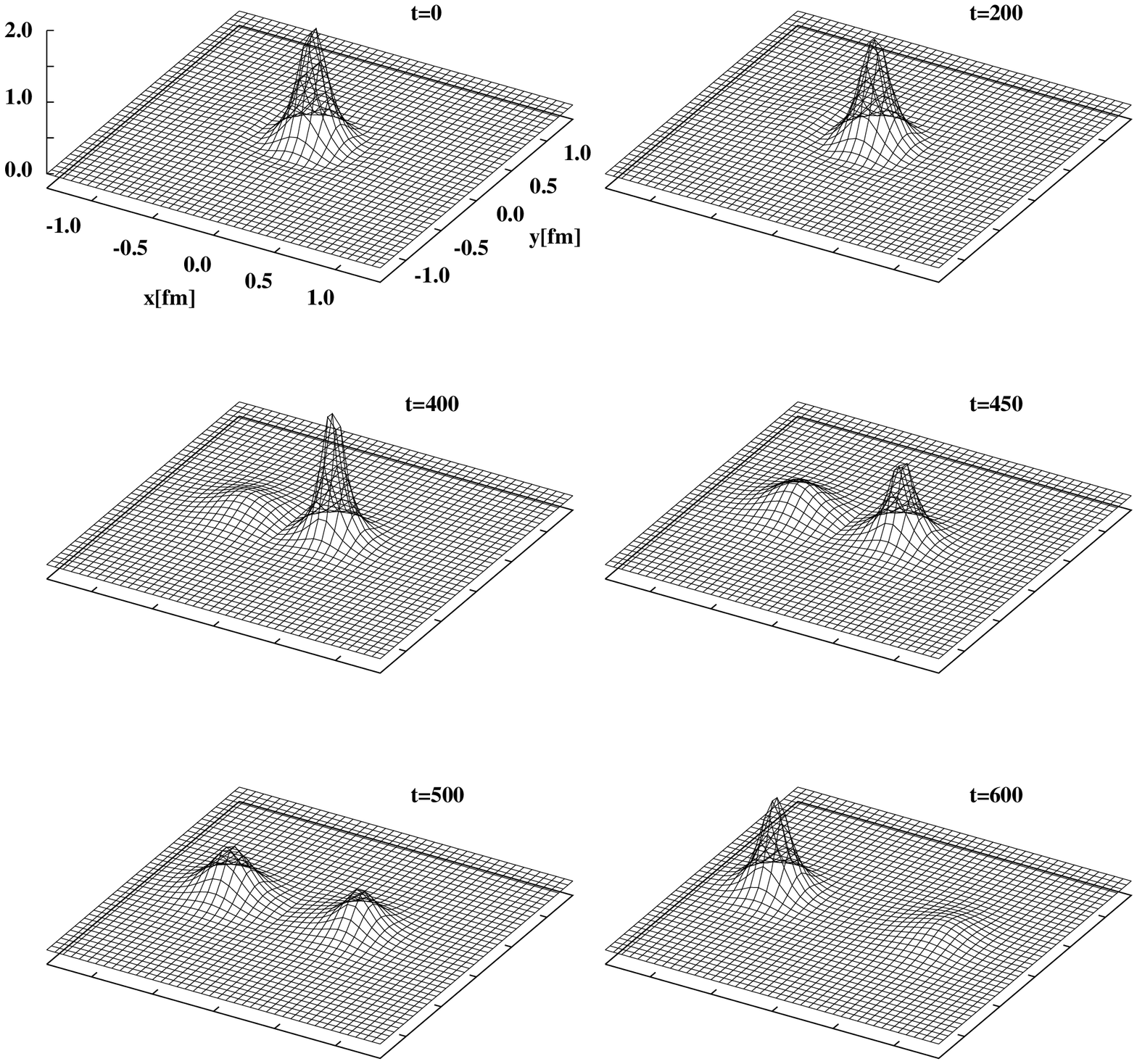}
{\small Fig.~7. The surface plot of the baryon number density for the
'twisting' mode.
\par}
\end{figure}
Instead of splitting in the middle, the fields are now twisting outwards.
This is understandable from the 'onion structure' of the baryon
density for the B=2 hedgehog. The density distribution is that of a
single spherically symmetric B=1 Skyrmion around which a second Skyrmion is
wrapped uniformly to cover the whole surface.
The inner and outer Skyrmions are initially rotated
relative to each other by the fluctuation field as can be
determined by using the symmetry relation for the $U$-field (\ref{sym}).
After the excitation, they start unwinding,
inside out, which leads to a twist of the full baryon number density. The
inner Skyrmion moves in one direction while the outer shell, also carrying
baryon number one, recedes into the opposite direction.
In addition, the solitons start rotating in space as well as in isospin space.
While the spatial rotation can be determined we were not able to reliably
calculate the isorotation due to lattice errors and the complicated nature
of the motion. As compared to the magnetic (or fission) mode now two angles
can vary.
The calculated energies and the relative separation coordinate $R$ for the
 magnetic path are shown in fig.~8.
\begin{figure}[h]
\vskip 7.5cm
\includegraphics{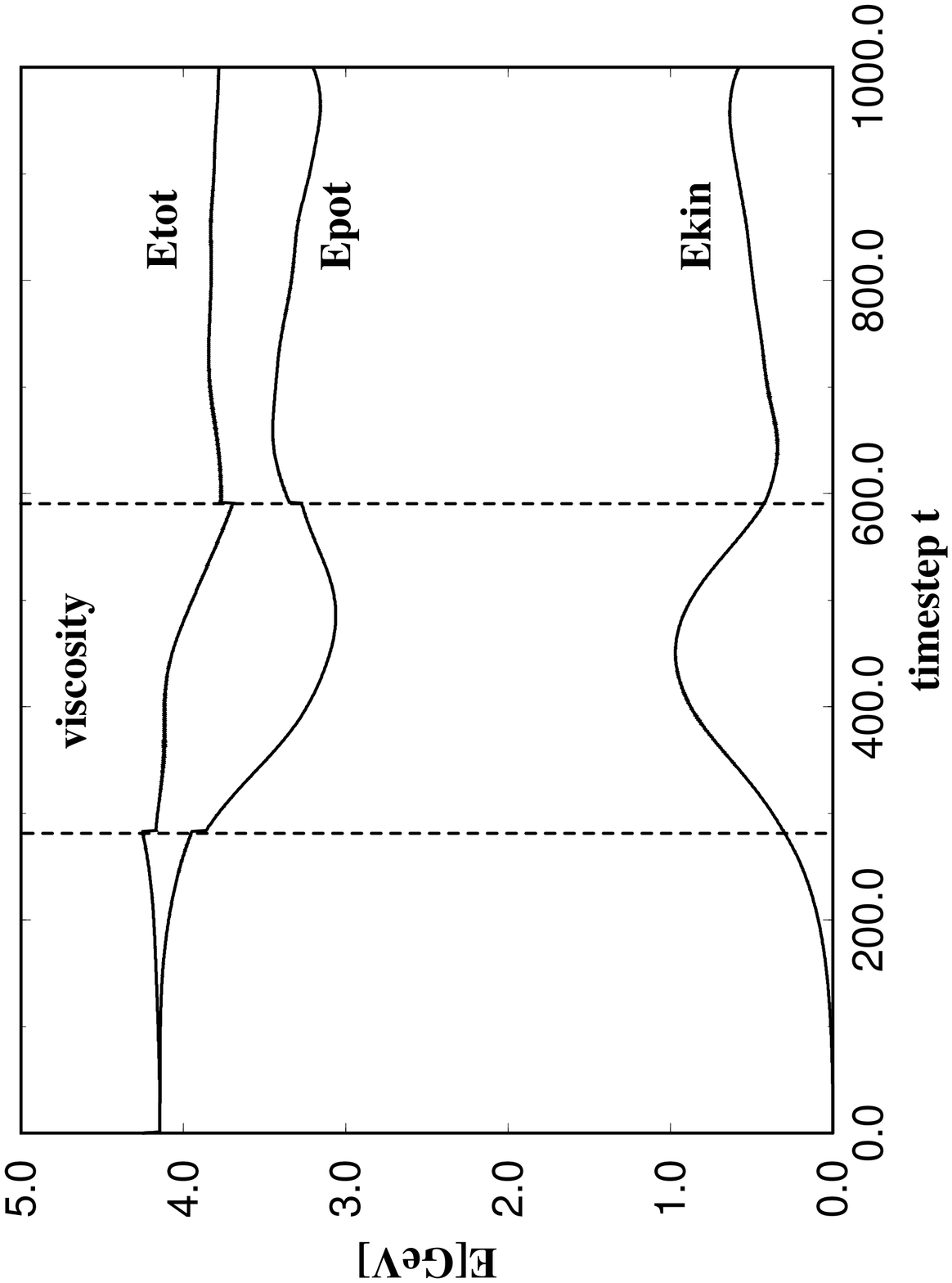}
\includegraphics{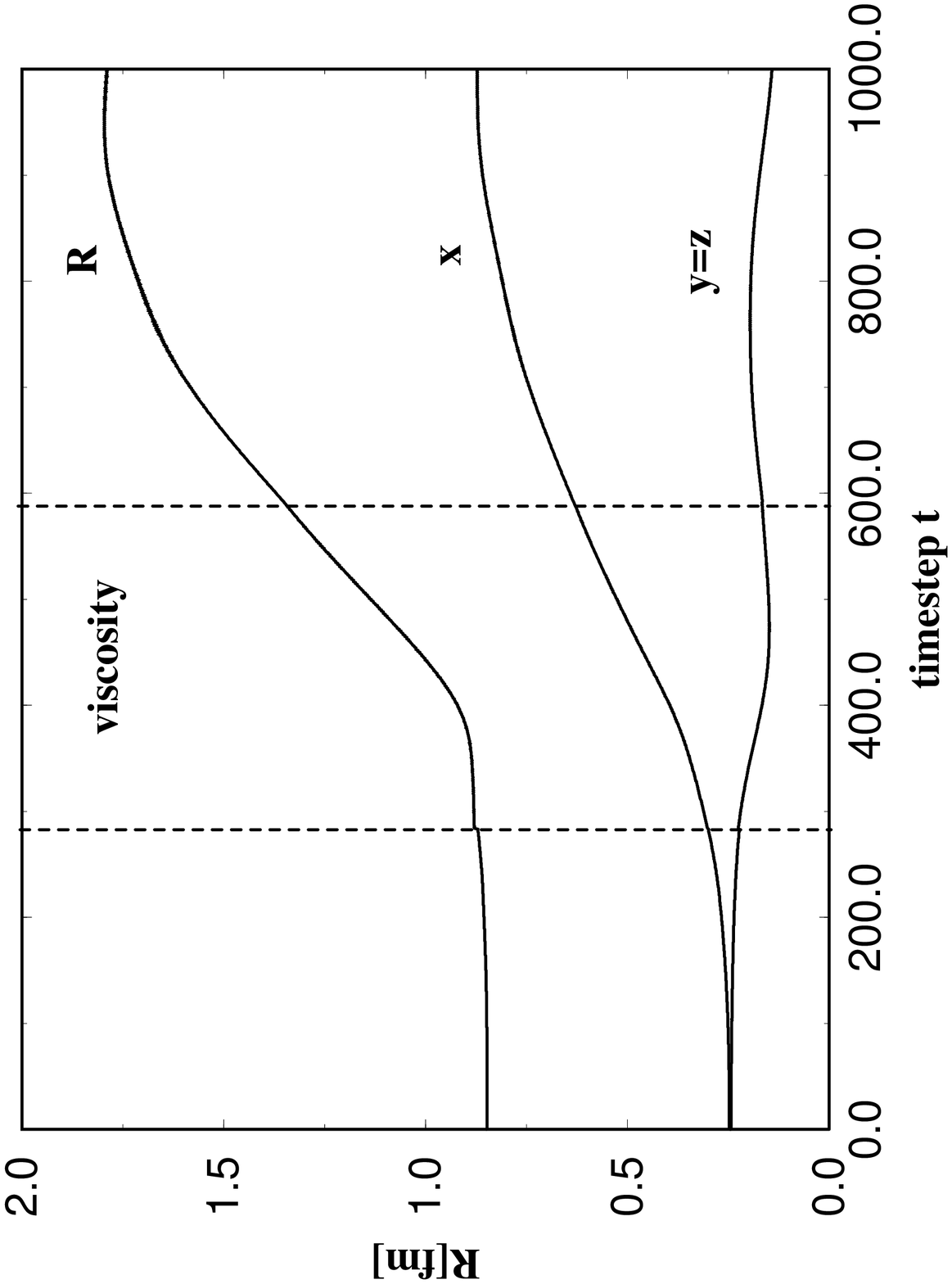}
{\small Fig. 8. The energies and  cartesian coordinates of relative separation
for the 'twisting' mode.
\par}
\end{figure}
The results are qualitatively similar to the magnetic mode. Note however that
it takes longer for the electric mode to start splitting since it is
less unstable than the magnetic mode.
The excitation of vibrational modes in the receding Skyrmions is
even more pronounced, leading to very large amplitudes in the baryon number
density, as it can be seen from fig.~7.
\subsection{The path of steepest descent}
The two paths studied in the previous sections have shown large excitations
of vibrational modes for the separating B=1 Skyrmions. This is due to the fact
that, by  evolving the full system in time, we never really can follow the
paths of steepest descent for the B=2 system. Only in the beginning of
the iteration, when the energy is small, are we sure that the path will be
close to the steepest one. For larger kinetic energies, however, we can easily
leave this path and the analysis in terms of zero or near zero modes becomes
exceedingly difficult due to the large number of vibrational modes.\\
To stay on paths of steepest descent, one can proceed in several ways.
One method is to readjust for each superposition of background field
and fluctuation the direction by calculating the new unstable mode for the
full configuration (\ref{fullconf}). This precisely would
define the path of steepest descent.
The other method is to evolve the fields to a given time and then minimize the
total energy at that instant of time by constraining the relative coordinates
of the B=2 system. In this way one can get rid of internal excitations of the
Skyrmions thus staying very close to the path of steepest descent.
Probably this is the method to be used in the future. \\

With large viscous damping during the entire time evolution we can find paths,
which should be close to some paths of steepest descent.
One nice result is shown in fig.~9 , where a combination of
magnetic and electric mode as initial fluctuation field for the excitation of
the B=2 hedgehog has been used.
\begin{figure}[h]
\vskip 10.5cm
\includegraphics{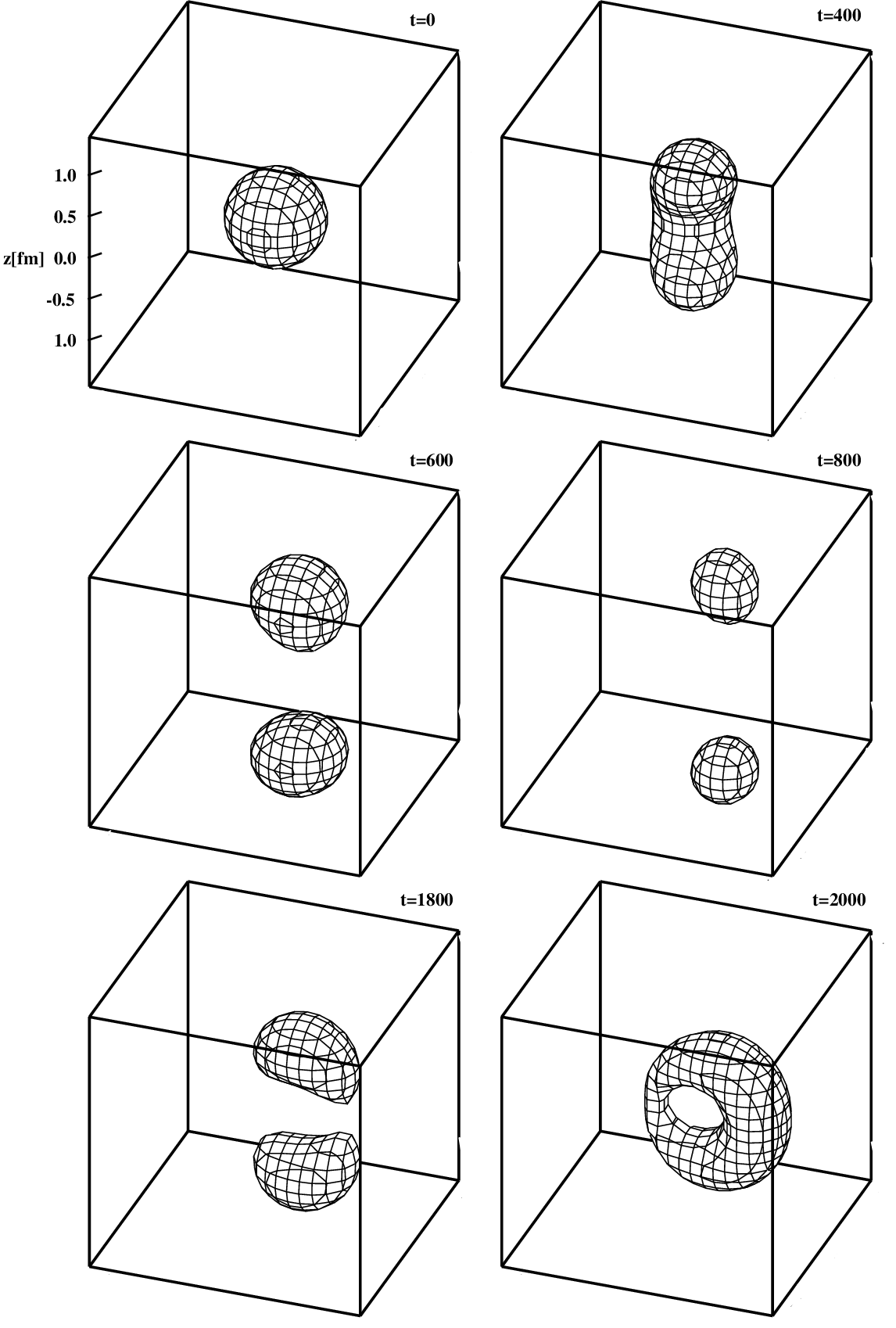}
{\small Fig.~9. Contour plot of the baryon number density for a fully
slowed down path starting from an initial excitation of the B=2 hedgehog
with a superposition of magnetic and electric modes.
\par}
\end{figure}
A specific mixture is necessary to orient the relative isospin such that the
path ends up in the donut configuration. The existence of this path has been
anticipated by
Manton \cite{Manton} and is confirmed by our calculations.
\section{Summary and Outlook}
We have performed a stability analysis of the B=2 hedgehog on the lattice by
solving the
equations of motion for the fluctuation fields via relaxation employing
the Leapfrog method in pseudotime. As conjectured by Manton \cite{Manton} the
hedgehog has six unstable modes, also on the lattice.
Following the excited B=2 hedgehog in real time we find essentially two
 different
decay pattern corresponding to two energetically different unstable and
three-fold degenerate modes. One is the fission mode, where the baryon number
density splits in the middle. This 'magnetic' mode has been found previously
on the $S^3$ sphere \cite{WiBa} The other type is the 'twister mode', where the
inner B=1 Skyrmion of the B=2 hedgehog is twisting outwards from the full
configuration. This mode is analogous to the 'electric' mode on the $S^3$
sphere \cite{WiBa}. Crossing the critical points of fission via viscous damping
we see a complicated motion in space and isospace. Besides the relative motion,
relevant for  semiclassical quantization in terms of zero and near zero modes,
the receding Skyrmions are excited vibrationally. The latter fact complicates
the situation drastically. To avoid the excitation of vibrational modes
one has to follow paths of steepest descent. As indicated in sect.~5.3 this
could be achieved by using a hybrid method of constrained
static and free time-dependent lattice calculations.
However, for this kind of method
one has to constrain the fields for various orientations
in order to calculate the metric of the collective coordinate manifold
via the derivatives of the fields with respect to
the coordinates (\ref{mass}).
This is probably the only way to extract the inertia tensor of the
B=2 Skyrmion system on the lattice.
\section*{Acknowledgments}{
We would like to thank Neil Snyderman for fruitful discussions.
This work was supported in part by the grant NSF PHY94-21309.
One of us (T.W.) acknowledges support from the German Academic Exchange
Service (DAAD) under the program HSPII/AUFE.}
\bibliographystyle{nhl}
\bibliography{B2hedge}

\end{document}